\documentclass[aps,pre,groupedaddress,showpacs,preprint]{revtex4}
\usepackage{graphicx}
\usepackage[dvips]{epsfig}
\usepackage{dcolumn}
\usepackage{subfigure}%
\usepackage{bm}
\usepackage{amssymb}
\usepackage{amsmath}

\begin{document}

\title{Two species coagulation approach to consensus by group level
interactions}
\author{Carlos Escudero$^\dag$, Fabricio Maci\`{a}$^\ddag$, and Juan J. L.
Vel\'{a}zquez$^\S $}
\affiliation{$\dag$ ICMAT (CSIC-UAM-UC3M-UCM), Departamento de Matem\'{a}ticas, Facultad
de Ciencias, Universidad Aut\'{o}noma de Madrid, Ciudad Universitaria de
Cantoblanco, 28049 Madrid, Spain \\
$\ddag$ Universidad Polit\'{e}cnica de Madrid, ETSI Navales, Avda. Arco de
la Victoria s/n, 28040 Madrid, Spain \\
$\S $ ICMAT (CSIC-UAM-UC3M-UCM), Facultad de Matem\'{a}ticas, Universidad
Complutense, 28040 Madrid, Spain}

\begin{abstract}
We explore the self-organization dynamics of a set of entities by
considering the interactions that affect the different subgroups conforming
the whole. To this end, we employ the widespread example of coagulation
kinetics, and characterize which interaction types lead to consensus
formation and which do not, as well as the corresponding different
macroscopic patterns. The crucial technical point is extending the usual one
species coagulation dynamics to the two species one. This is achieved by
means of introducing explicitly solvable kernels which have a clear physical
meaning. The corresponding solutions are calculated in the long time limit,
in which consensus may or may not be reached. The lack of consensus is
characterized by means of scaling limits of the solutions. The possible
applications of our results to some topics in which consensus reaching is
fundamental, like collective animal motion and opinion spreading dynamics,
are also outlined.
\end{abstract}

\maketitle

\section{Introduction}

The property that characterizes those systems which long range
order is due to the coupling of simple interactions at their
different scales is usually denoted as emergence. It has been
studied in many different contexts, ranging from collective animal
motion \cite{couzin} to linguistic dynamics \cite{cucker} and
passing through condensed matter \cite{escudero} and financial
systems \cite{lux}. It is thus a fundamental property of many
important systems which encodes essential information about them.
Emergence is commonly studied as arising directly from the
interactions of the most fundamental units configuring the system.
However, self-organization does not necessarily happen this way.
Macroscopic order may be a consequence of interactions at
different scales, in which the diverse subsystems play a
particular contributing role. We will illustrate how this occurs
by means of studying the dynamics of particular coagulation
equations. While the origins and consequences of the arising of
consensus by group level interactions go far beyond the use of
coagulation equations, these constitute a reasonable approach to
the subject, precisely due to their universality. Indeed,
coagulation equations have proved their utility in topics as
diverse as aerosols \cite{seinfeld}, polymerization \cite{ziff},
Ostwald ripening \cite{ls,meerson}, galaxies and stars clustering
\cite{silk}, and population biology \cite{niwa} among many others.
They thus exemplify the broad applicability that the understanding
of this sort of ordering processes may bring.

A classical example of self-organizing systems is collective
organism behavior, which has been studied using a broad range of
theoretical techniques \cite{bertozzi,bodnar,orsogna}. This
includes the adaptation of some of the classical models of
statistical mechanics of spin systems~\cite{vicsek}. Indeed,
statistical physics models and methods have been successfully
adapted to situations of a completely different nature such as
opinion formation and spreading~\cite{toral,fortunato}. One of the
classical models in this context is the voter
model~\cite{voter1,voter2,liggett}, which has been studied within
both fields of statistical mechanics and probability theory. Due
to the large generality in which these and similar processes
appear it seems appealing extending coagulation models to two
species dynamics. On one hand, the extension is natural from a
methodological viewpoint; on the other hand, it smoothly matches
with previous approaches in the subject of self-organization.
Clustering has been previously studied in population
dynamics models~\cite{emilio} including swarming systems~\cite{clusterswarm}%
, and coagulation equations have been used in both swarming~\cite{coaguswarm}
and opinion formation models~\cite{toral2}. We shall show in the following
how two-species coagulation may constitute a sensible extension of previous
approaches.

There exist many biological examples of collective motion,
including insects, birds, fishes... In these cases the interaction
among individuals gives rise to a consensus which consists in the
selection of the direction of motion. A specific example that has
been extensively studied both from the experimental and
theoretical point of view is the problem of locusts swarming. The
experiment performed in \cite{buhl} revealed that locusts marching
on a (quasi one dimensional) ring presented a coherent collective
motion for high densities; low densities were characterized by a
random behavior of the individuals and intermediate densities
showed coherent displacements alternating with sudden changes of
direction. The models that have been introduced to describe this
experiment assume that the organisms behave like interacting
particles~\cite{buhl,yates}. Related interacting particle models
have been used to describe the collective behavior of many
different organisms and analyzing the mathematical properties of
such models is currently a very active research
area~\cite{bertozzi,orsogna,vicsek}.

In this paper we will not try to describe the detailed behavior of
any specific organism. Our goal will be to explore the
mathematical consequences of models of organism clustering, that
is an alternative way of describing the collective behavior of
these systems. This approach does not rule out the possibility of
particle-like interactions among the components of the system, but
it assumes that the tendency to follow the behavior of the
neighborhoods is strong enough to produce collective effects at
the level of the whole cluster instead of individual elements. As
we shall see, this experiment suggests interesting two-species
coagulation dynamics. We also expect that our approach may be of
some relevance in the field of opinion formation, again as an
idealized first approximation describing the dynamics of two
opinions. We will present a simplistic model which otherwise has a
number of interesting features. It has a number of simplifying
assumptions: 1) only two directions of motion are considered,
while in realistic cases there are infinite; 2) it is assumed that
interactions take place at the group level and pair interactions
are not taken into account; 3) the probability laws of interaction
are very specific and simple in order to allow for mathematical
tractability. Despite its simplicity, our model is still able to
predict consensus formation in some cases and to give rise to
counterintuitive results. Coagulation models have been also used
to describe collective organism motion in~\cite{coaguswarm}.

We start considering a one dimensional spatial situation in which the
particles move in clusters of $\ell$ individuals and $f^+(\ell,t)$ ($%
f^-(\ell,t)$) represents the spatial density of clusters moving towards the
right (left) in the time slot $[t,t+dt)$. Clusters are modified when they
collide with other clusters, in such a way that the probability distribution
obeys the following equation of motion
\begin{equation}  \label{smoluchowsky}
\partial_t f^\mp(\ell,t) = \sum_{m,k,j} \left[ \Phi(k,j;m,\ell)
f^\pm(k,t)f^\mp(j,t)- \Phi(m,\ell;k,j) f^\pm(m,t) f^\mp (\ell,t) \right],
\end{equation}
where $\Phi(k,j;m,\ell)$ is the collision kernel: it states the Poissonian
probability rates with which a collision among clusters with $k$ and $j$
particles occurs and yields clusters with $m$ and $\ell$ particles. This
equation has been derived under the assumption that the clusters interact
rarely with colliding clusters (i. e., the interaction Poissonian rates are
small), in such a way that a well stirred distribution of clusters results.
Therefore, it will be assumed that the distribution of cluster sizes is
uncorrelated. Then the Boltzmann Stosszahlansatz $f_2^\pm(\ell,k,t)=f_1^\pm(%
\ell,t) f_1^\pm(k,t)$ follows, where $f_2$ and $f_1$ represent the
two-clusters and one-cluster distribution function respectively
(the subscript $1$ has been omitted in Eq. (\ref{smoluchowsky})).
Indeed, note that assuming small Poissonian rates physically
indicates that only in a small fraction of the collisions there is
a successful interaction among the clusters. We additionally
assume that the collision kernel is symmetric under reflections
$\Phi(k,j;m,\ell)=\Phi(j,k;\ell,m)$ as a consequence of spatial
isotropy. A characteristic property of this sort of kernels is
that all particles form a single cluster after a successful
collision occurs. Note that Eq. (\ref{smoluchowsky}) is a
two-species coagulation equation. The mathematical structure of
one-species coagulation equations with exactly solvable kernels is
relatively well understood (see for instance
\cite{mcleod,penrose,menon}), and we will build our progress on
the unexplored two-species coagulation based on these previous
works.

\section{Random kernel}

To study Eq. (\ref{smoluchowsky}) we introduce the generating function
\begin{equation}
F^\pm (z,t)=\sum_{\ell=0}^\infty f^\pm(\ell,t) z^\ell,
\end{equation}
and we will consider a family of kernels of an apparent physical meaning.
Our first choice is
\begin{eqnarray}  \label{collisionkernel1}
\Phi(\ell,m;0,m+\ell) &=& \Phi(\ell,m;m+\ell,0)=1/2, \\
\Phi(k,j;\ell,m) &=& 0, \qquad \mathrm{otherwise},  \label{collisionkernel2}
\end{eqnarray}
meaning that when a collision takes place the colliding clusters merge into
a single one which chooses its direction of motion with equal probability.
The generating function allows to transform the coagulation equations into
the differential equations system
\begin{eqnarray}  \label{ztransform1}
\partial_t F^+(z,t) &=& \frac{1}{2} F^+(z,t)F^-(z,t)-F^+(z,t)F^-(1,t), \\
\partial_t F^-(z,t) &=& \frac{1}{2} F^+(z,t)F^-(z,t) - F^-(z,t)F^+(1,t).
\label{ztransform2}
\end{eqnarray}
The number of clusters moving in each direction will be denoted as
\begin{equation}
N^\pm(t)=F^\pm(1,t),
\end{equation}
and by employing Eq. (\ref{ztransform1}) and Eq. (\ref{ztransform2}) we find
\begin{eqnarray}  \label{difference}
N^+(t)-N^-(t)=C_0, \\
\partial_t N^-=-\frac{1}{2}N^-(C_0+N^-),  \label{logistic}
\end{eqnarray}
for a constant $C_0$. Physically Eq. (\ref{difference}) means that in each
successful interaction between clusters a cluster is eliminated and, on
average, the same number of clusters moving towards the right and towards
the left are eliminated. Eq. (\ref{logistic}) can be integrated to yield
\begin{equation}
N^-(t)=\frac{C_0 N^-(0)}{e^{C_0 t/2}N^+(0)-N^-(0)},
\end{equation}
what reveals that $N^-(t) \to 0$ and $N^+(t) \to C_0$ if $N^+(0)>N^-(0)$; if
$N^-(0)>N^+(0)$ then $N^+(t) \to 0$ and $N^-(t) \to -C_0$. The number of
particles moving in each direction will be denoted as
\begin{equation}
M^\pm(t)=\partial_z F^\pm(1,t).
\end{equation}
By differentiating Eqs. (\ref{ztransform1}) and (\ref{ztransform2}) we find
the differential system
\begin{eqnarray}  \label{mean1}
\partial_t M^+ &=& \frac{1}{2} M^-N^+ - \frac{1}{2} M^+N^-, \\
\partial_t M^- &=& \frac{1}{2} M^+N^- - \frac{1}{2} M^-N^+,  \label{mean2}
\end{eqnarray}
which translates into the equations
\begin{eqnarray}
M^+ + M^- = C_1, \\
\partial_t M^- = \frac{1}{2} N^- (C_1-M^-)-\frac{1}{2}M^-N^+,
\end{eqnarray}
where $C_1$ is a constant. The first equation expresses the
physical fact that the total number of particles is conserved. The
second equation is a linear ordinary differential equation which,
as we know the values of $N^+$ and $N^-$, can be exactly
integrated by means of the method of variation of parameters. Its
solution reads
\begin{equation}
M^-(t)=\frac{2C_1N^-(0)^2 + e^{C_0t/2} [2C_0^2 M^-(0)-2C_1N^-(0)^2+ C_0 C_1
N^-(0) N^+(0)t]} {2[N^-(0)-e^{C_0t/2}N^+(0)]^2},
\end{equation}
which indicates us that $M^-(t) \to 0$ and $M^+(t) \to C_1$ if $N^+(0)>N^-(0)
$; if $N^-(0)>N^+(0)$ then $M^+(t) \to 0$ and $M^-(t) \to C_1$. It is clear
that the particles interacting as prescribed by the equidistributing kernel (%
\ref{collisionkernel1})--(\ref{collisionkernel2}) merge into $C_0$
separate groups travelling in the same direction in the long time
limit. In other words, we say that consensus is reached and that
the corresponding macroscopic pattern is an ordered one with all
particles travelling in the same direction. Another characteristic
of this type of evolution, as we have seen, is the small variation
of the cluster sizes during the transient toward consensus.

The degenerated case in which $N^+(0)=N^-(0)$ behaves differently. Now the
solution adopts the universal self-similar form
\begin{equation}  \label{asymptotic}
f^{\pm}(\ell,t)= \frac{8}{M t^2} \exp \left( -\frac{4\ell}{M t} \right)
\sum_{m=1}^\infty \delta_{md,\ell} + o \left( \frac{1}{t} \right),
\end{equation}
uniformly in $\ell$, where $M=(M^+ + M^-)/2$ is a constant of the dynamics
and $d= \mathrm{gcd} \{ n | a_n^\pm \neq 0 \}$, where $F^\pm(z,0)=%
\sum_{n=0}^\infty a_n^\pm z^n$. This solution develops in the limit $\ell
\gg d$, $t \gg 4d/M$ and $t \approx \ell$. This self-similar asymptotic
result can easily derived assuming $F^+(z,0)=F^-(z,0)$, which in turn
implies $F^+(z,t)=F^-(z,t)$ for all times. In this case Eqs.~(\ref%
{ztransform1})-(\ref{ztransform2}) can be exactly integrated,
because they reduce to a single equation of the logistic type.
Using this solution into the Cauchy formula
\begin{equation}  \label{cauchy}
f^\pm (\ell,t)=\frac{1}{2 \pi i} \int_{|z|=1} \frac{F^\pm(z,t)}{z^{\ell+1}}%
dz,
\end{equation}
where the line integration is carried out in the counterclockwise direction,
one recovers the cluster density. Taking the long time limit on Eq.~(\ref%
{cauchy}) one easily finds Eq.~(\ref{asymptotic}) after a formal
calculation. A rigorous proof of this asymptotic result is possible even
when we have just the equality $N^+(0)=N^-(0)$, and $F^+(z,0)$ and $F^-(z,0)$
differ otherwise. However the proof is long and technical, and so it will be
reported elsewhere \cite{emv}. The physical consequence is on the other hand
straightforward, and it reveals that consensus is not reached in this case.
Contrarily, clusters with higher numbers of components are being generated
at every moment in both directions, and this process continues for all
times. Simultaneously, clusters with lower numbers of components tend to
disappear as time evolves. Both characteristics can be read from the
asymptotic self-similar form Eq.~(\ref{asymptotic}). With respect to the
number of particles propagating in each direction $M^\pm(t)$ they tend to
equilibrate each other, and they stay the same for all times if they are
equal initially. This is a consequence of the dynamics given by Eqs.~(\ref%
{mean1})-(\ref{mean2}) when the number of clusters travelling in
each direction is the same. The physical reason underlying this
behavior is the cluster availability to cause collisions for
arbitrarily long times, and this is precisely what produces an
unbounded cluster growth. Alternatively, the dynamics of the
random kernel can be understood in terms of the initial difference
among clusters moving in the different directions $C_0$. As stated
in Eq.~(\ref{difference}) this difference is conserved for all
times, and it becomes more visible as the annihilation of
clusters, mediated by successful interactions, progresses. The
final outcome is a set of $C_0$ clusters travelling in the
direction in which the majority of clusters was travelling
initially.

\section{Majority kernel}

We now consider a different kernel which is still integrable. Let us propose
the following coagulation equations
\begin{equation}  \label{collisionkernel3}
\partial_t f^\mp(\ell,t) = \sum_{k+j=\ell} \frac{k}{k+j} f^\pm(k,t)f^%
\mp(j,t)- \sum_{m} f^\pm(m,t) f^\mp (\ell,t).
\end{equation}
In this case two colliding groups merge again into a single one, which
direction of motion is this time biased. It is more likely that the
generated group moves in the same direction as the most populated colliding
cluster. Specifically, if a group of $k$ particles collides with a group of $%
j$ particles there is a probability $k/(k+j)$ that the resulting group
continues to travel in the same direction and a probability $j/(k+j)$ that
the direction shifts. This interaction rule apparently constitutes a
stronger trend toward order than the equidistributing one, but however its
consequences are rather unexpected as we will see. Using the generating
function formalism as in the former case we find
\begin{eqnarray}
\partial_t (\partial_z F^+) &=& F^- (\partial_z F^+) - (\partial_z F^+) N^-,
\\
\partial_t (\partial_z F^-) &=& F^+ (\partial_z F^-) - (\partial_z F^-) N^+,
\end{eqnarray}
where the number of clusters $N^\pm (t)= F^\pm(1,t)$. For the number of
particles $M^\pm (t)=\partial_z F^\pm(1,t)$ we find
\begin{equation}
\partial_t M^+=\partial_t M^-=0.
\end{equation}
In consequence, the number of particles travelling in each
direction stays constant for all times. In this case consensus is
not reached, and the macroscopic pattern corresponds identically
to the initial disordered state for all times. This result
contrasts with the former one, in which case an apparently weaker
interaction rule leaded to order for long times. Physically, this
effect takes place because during most of the successful
interactions between particles, the majority forces a change in
the direction of motion of the particles moving in the opposite
direction. However, the rare collisions in which the minority wins
produces a change in the direction of motion of several particles.
At the end, the balance between these two effects implies that on
average the total number of particles moving in each direction
remains constant.

In this second case we can gain further insight into the dynamics by
proposing the solution ansatz
\begin{equation}  \label{laplacess}
F^\pm(z,t)=\frac{1}{t}\varphi^\pm [(z-1)t],
\end{equation}
which corresponds to the real space universal self-similar form
\begin{equation}
f^{\pm}(\ell,t)= \frac{1}{t^2} \mathcal{G}^\pm \left( \frac{\ell}{t} \right)
\sum_{m=1}^\infty \delta_{md,\ell} + o \left( \frac{1}{t} \right),
\label{selfsimilar}
\end{equation}
where $\mathcal{G}^\pm$ is the scaling function, in clear correspondence
with the asymptotic solution of the equidistributing dynamics with the
degenerated initial condition, see Eq.~(\ref{asymptotic}). Instead of giving
a rigorous proof of this fact we offer here numerical evidence of it.

To numerically compute the self-similar solution we consider the continuum
equation
\begin{equation}  \label{continuum}
\partial_t f^\pm (x,t)= \int_0^x \frac{y}{x} f^\pm(y,t)f^\mp (x-y,t)dy
-f^\pm(x,t) \int_0^\infty f^\mp (y,t)dy.
\end{equation}
By analogy with the classical coagulation equations it is reasonable to
expect that the behavior of the discrete equation will approach the behavior
of integral continuous equation as $\ell/t$ becomes smaller. The Laplace
transform of its solution is
\begin{equation}
\hat{f}^\pm (z,t)=\int_0^\infty f^\pm (x,t) e^{-zx} dx,
\end{equation}
which is defined for $\mathrm{Re}(z)>0$. We assume the self-similar form in
Laplace space
\begin{equation}  \label{laplace2}
\hat{f}^\pm (z,t)=\frac{1}{t} \varphi^\pm (zt)=\frac{1}{t} \varphi^\pm (\xi),
\end{equation}
where $\xi=zt$ is the self-similar variable and $\mathrm{Re}(\xi)>0$. This
corresponds to the self-similar form in real space
\begin{equation}  \label{realspacess}
f^\pm (x,t)=\frac{1}{t^2} \mathcal{G}^\pm \left( \frac{x}{t} \right),
\end{equation}
where the self-similar profile can be recovered from the inverse Laplace
transform
\begin{equation}  \label{bromwich}
\mathcal{G}^\pm (\zeta)=\frac{1}{2 \pi i} \int_{\gamma -i\infty}^{\gamma
+i\infty} e^{\zeta \eta} \varphi^\pm (\eta) d\eta,
\end{equation}
where $\gamma \in \mathbb{R}$ is large enough (larger than the real part of
all the poles of $\varphi^\pm(\cdot)$) and we have used the Bromwich
integral formula~\cite{arfken}. We note the analogy of these formulas with
Eqs.~(\ref{asymptotic})-(\ref{cauchy}) and Eqs.~(\ref{laplacess})-(\ref%
{selfsimilar}), i. e., we are again calculating the scaling form of the
solution in the Laplace and real space formulations. The scaling functions
obey the system
\begin{eqnarray}
\xi \partial_{\xi \xi} \varphi^+ &=& [\varphi^- -\varphi^-(0)]
\partial_{\xi} \varphi^+, \\
\xi \partial_{\xi \xi} \varphi^- &=& [\varphi^+ -\varphi^+(0)]
\partial_{\xi} \varphi^-,
\end{eqnarray}
as can be seen by making the substitution~(\ref{laplace2}) directly into
Eq.~(\ref{continuum}). By means of the new substitutions $W^+=\varphi^+
-\varphi^+(0)$ and $W^-=\varphi^- -\varphi^-(0)$ we arrive at the system of
ordinary differential equations:
\begin{equation}
\xi W^+_{\xi \xi} = W^- W^+_\xi, \qquad \xi W^-_{\xi \xi} = W^+ W^-_\xi.
\end{equation}
Changing variables $\xi=-e^\tau$ and $\psi^\pm=W^\pm+1$ we arrive at the
four-dimensional dynamical system
\begin{eqnarray}
\label{h1}
H^+_{\tau} &=& \psi^- H^+, \\
\label{psi1}
\psi^+_\tau &=& H^+, \\
\label{h2}
H^-_{\tau} &=& \psi^+ H^-, \\
\label{psi2}
\psi^-_\tau &=& H^-,
\end{eqnarray}
where Eqs.~(\ref{psi1}) and (\ref{psi2}) are actually the
definitions of $H^+$ and $H^-$ respectively. This system is
subject to the initial conditions
\begin{equation}
\psi^\pm(-\infty)=1, \qquad H^\pm(- \infty)=0,
\end{equation}
and which long time behavior is
\begin{equation}
H^\pm(+ \infty)=0, \qquad \psi^\pm(+\infty)=1-\varphi^\pm(0),
\end{equation}
as can be deduced from the different changes of variables that have been
performed so far. This dynamical system defines the self-similar solutions~(%
\ref{laplace2}) to coagulation equation~(\ref{continuum}). So we devote the
rest of this section to showing that these solutions actually exist. In the
dynamical system Eqs. (\ref{h1})-(\ref{psi1})-(\ref{h2})-(\ref{psi2}) we can
identify the invariant hyperplanes $\{H^\pm = 0\}$ and the invariant plane $%
\{\psi^+=\psi^-, \, H^+=H^-\}$. All the points belonging to the plane $%
\{H^+=H^-=0\}$ are fixed points; and all the fixed points of this system
belong to this plane. The quantity $E=\psi^+\psi^- -H^+ -H^-$ is a first
integral of motion, and for the initial conditions under consideration $E=1$%
. This means that the trajectories under consideration, which depart from
the plane $\{H^+=H^-=0\}$ and go back to this plane in the infinite time
limit are distributed in the hyperbola $\psi^+ \psi^-=1$ precisely in this
limit.

\begin{figure}[tbp]
\begin{center}
\psfig{file=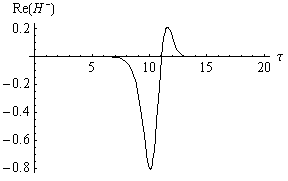,width=7.5cm,angle=0}
\end{center}
\caption{Numerical solution $\mathrm{Re}(H^-)$ versus $\protect\tau$ of
system Eqs. (\protect\ref{h1}), (\protect\ref{psi1}), (\protect\ref{h2}),
and (\protect\ref{psi2}) integrated in the complex plane. The initial
conditions are $\mathrm{Re}(H^+)=1-\mathrm{Re}(\protect\psi^+)=-10^{-4}\cos(%
\protect\theta_1)\cos(\protect\theta_2)$, $\mathrm{Re}(H^-)=1-\mathrm{Re}(%
\protect\psi^-)=-10^{-4}\sin(\protect\theta_1)\cos(\protect\theta_2)$, $%
\mathrm{Im}(H^+)=\mathrm{Im}(\protect\psi^+)=-10^{-4}\cos(\protect\theta%
_1)\sin(\protect\theta_2)$, $\mathrm{Im}(H^-)=\mathrm{Im}(\protect\psi%
^-)=-10^{-4}\sin(\protect\theta_1)\sin(\protect\theta_2)$, $\protect\theta%
_1=39\protect\pi/200$, $\protect\theta_2=\protect\pi/2$.}
\label{numerics}
\end{figure}

We have numerically integrated system Eqs. (\ref{h1}), (\ref{psi1}), (\ref%
{h2}), and (\ref{psi2}); for an example see Fig. \ref{numerics}. These
trajectories numerically define the self-similar profiles $\varphi^\pm(\cdot)
$. We still have to check that the inverse Laplace transform formula~(\ref%
{bromwich}) makes sense for these numerical solutions. It turns out that the
functions $\varphi^\pm(\cdot)$ (considered as functions of the variable $\tau
$, which is now considered to be complex) can be analytically extended to
the regions $\mathrm{Re}(\tau)>L$, $\mathrm{Re}(\tau)<1/L$, $|\mathrm{Im}%
(\tau)|<\pi/2+\epsilon$ for some real $L$ sufficiently large and some real $%
\epsilon>0$ sufficiently small (see \cite{emv}). On the other hand, we have
checked numerically that in fact there are no singularities for the
functions $\varphi^\pm(\cdot)$ in the whole strip $|\mathrm{Im}%
(\tau)|<\pi/2+\epsilon$, $\mathrm{Re}(\tau)\in(-\infty, \infty)$. This
implies that the functions $\varphi^\pm(\cdot)$ considered as functions of $%
\xi$ (considered also as a complex variable) are analytic in the half-plane $%
\mathrm{Re}(\xi)>0$ and therefore the function $\mathcal{G}^\pm (\cdot)$ can
be obtained by means of the inversion formula for the Laplace transform (\ref%
{bromwich}). This numerically proves the existence of scaling solutions to
coagulation Eq.~(\ref{continuum}) obeying the self-similar scaling~(\ref%
{realspacess}).

\section{Conclusions}

To conclude, the emergent properties of physical systems are
determined by the interactions among its fundamental constituents,
but they could also be described and understood in terms of the
interactions among the different subgroups conforming the whole.
Consensus is the microscopic characteristic underlying macroscopic
order in these systems, and it is reached or not depending on the
interaction nature. In the specific case of the majority kernel
considered in this paper, Eq.~(\ref{collisionkernel3}), the
absence of consensus is due to the finely tuned balance between
the transition probabilities of the kernel and the number of
elements changing their state in each interaction.
Kernel~(\ref{collisionkernel3}) can be easily generalized to
models with the form:
\begin{equation}
\partial _{t}f^{\mp }(l,t)=\sum_{k+j=l}\left( \frac{k}{k+j}\right) ^{\alpha
}f^{\pm }(k,t)f^{\mp }(j,t)-\sum_{m}f^{\pm }(m,t)f^{\mp }(l,t),
\qquad \alpha >0. \label{collisionkernel4}
\end{equation}
For this class of equations the finely tuned balance that takes
place for $\alpha =1$ does not take place in general. An
interesting mathematical question would be determining the range
of values of $\alpha$ yielding consensus formation for these
models. Notice that Eqs.~(\ref{collisionkernel1}) and
(\ref{collisionkernel2}) constitute an example of interaction
kernel which does not preserve the balances among collisions, and
thus leads to ordered long-range states, despite the seemingly
weaker trend toward order implied by it when compared to the one
in Eq. (\ref{collisionkernel4}).

In the simple cases we have studied, consensus takes the form of
one or several clusters travelling in the same direction. As two
directions are possible, and the system is symmetric with respect
to them, we note the similarity of this process with those systems
with two symmetric absorbing states~\cite{munoz,lopez}. It is
clear that consensus is an absorbing state in our model, as
collisions stop when consensus is reached. Complementarily, the
active state is the one in which collisions still take place,
which necessarily implies the existence of clusters travelling in
both directions and the concomitant absence of consensus. We
additionally note that for the dynamics given by the majority
kernel described in the last section, consensus is not reached in
the coagulation equation approximation. So this disordered state
could be thought of as a long lived, deterministically stable,
metastable state which will finally decay to one of the absorbing
states due to the effects of fluctuations. We have of course not
studied such effects in the present paper, so we simply mention
the possible similarity of this process with the dynamics of
certain systems characterized by the proximity to absorbing
states. Establishing a rigorous connection is far beyond the scope
of this work.

In the light of our results it seems possible that some self-organizing
systems interact according to some rule which does not preserve the balance
among the collisions or interactions at their different scales. Otherwise,
as we have seen, if some moment is preserved then consensus is consequently
not reached. As we have mentioned, these techniques may be applied or
adapted to swarming and opinion formation systems. Our approach constitutes
an idealized first approximation to these sorts of problems. More realistic
approaches should deal with the effect of fluctuations and finite sizes and
this way estimate how stochastic forces may influence ordering/disordering
processes. Also, we have not taken into account the effect of saturation and
fragmentation of clusters, which would keep the cluster masses finite. We
have focused for simplicity on the short time behavior during which the
initial growth of cluster masses could be described with the probabilistic
rules we have considered herein. We leave for future work the implementation
of these improvements.

\section*{Acknowledgments}

This work has been partially supported by the DGES Grants MTM2007-61755 and
MTM2008-03754. JJLV thanks Universidad Complutense for its hospitality.

\end{document}